\magnification=\magstep1
\input epsf.sty

\def\Referencja#1{\noindent\hangindent=50pt\hangafter=1\nobreak
\frenchspacing #1 \par}

\centerline{\bf Spectral Components in the Optical Emission of the Seyfert 
Galaxy NGC 5548}
\medskip
\centerline{\bf and the Comparison of Intrinsic Nuclear Spectra 
with Accreting Corona Model}

\bigskip

\centerline{by}

\bigskip

\centerline{J. Kuraszkiewicz, Z. Loska, and B. Czerny}

\centerline{Copernicus Astronomical Center, Bartycka 18, 00-716 Warsaw, Poland}

\centerline{e-mail:(jcuk,zbylo,bcz)@camk.edu.pl}

\bigskip\bigskip

\centerline{ABSTRACT}
\bigskip

We study the extensively monitored Seyfert galaxy NGC~5548 and compare
its nuclear emission with models of accretion disk with accreting 
corona. To obtain the intrinsic nuclear spectra from the observed spectra
we had to estimate and subtract 
the contribution from circumnuclear components 
such as stars, the Balmer continuum and blended FeII lines,
and the FC2 extended, featureless continuum. The true nuclear spectra
were compared with a two parameter model of the accreting disk with an
accreting corona, described by the mass of the central black hole and
viscosity. The model that best fitted the data was for 
$M_{BH}=1.4 \times 10^8 M_{\odot}$ and the viscosity parameter 
$\alpha=0.033$. Such a low viscosity parameter was necessary to
produce the sufficient amount of X-rays. The vertical outflow of mass
from corona in the form of wind had to be neglected in our model in order
to fit into high and low states that NGC~5548 underwent. The model also
predicts the behavior of the overall opt/UV/X continuum of NGC~5548
during the whole five year campaign.

\bigskip


\bigskip

\centerline{\bf 1. Introduction}
\bigskip

The Seyfert 1 galaxy NGC 5548 (z=0.0174) is one of a few active galaxies 
extensively
as well as systematically monitored over the last ten years as its nucleus
is strongly variable, bright and the galaxy is relatively nearby. 
The amount of
data collected for NGC 5548 actually exceeds all the other monitored AGN:
NGC~3783 (Reichert et al. 1994, Stripe 1994), Mrk 335 (Zheng et al. 1995)
or the BL Lac object PKS 2155-304 (Edelson et al. 1995). 	

The galaxy has been frequently observed since late seventies 
in the optical band, in
the UV with  IUE, and by a number of X-ray satellites. In 1988
a systematic campaign has been started by AGN Watch team. Optical 
spectra were taken by ground based telescopes every $\sim 4$ days since 
December 1988 (Peterson et al. 1991, 1992, 1994). UV variability was 
followed with IUE during first eight months of the campaign
(December 1988 - August 1989) with the same frequency sampling (Clavel et al.
1991). The X-rays and UV 
were measured simultaneously by IUE and Ginga satellite 11 times 
(Clavel et al. 1992). In 1993 the galaxy was monitored by HST and IUE
 in the UV band
(Korista et al. 1995). Soft X-ray variability was monitored by EUVE
satellite (Marshall, Fruscione \& Carone 1995) as well as ROSAT (e.g. Done
et al. 1995).

The collected data have been used for a number of studies of relative delays
in various energy bands (optical vs. UV, X-rays vs. opt and UV etc.)
as well as between the continuum and the broad emission lines. These data
offer also an unprecedented possibility to study  the variability of the
nuclear continuum thus offering a chance to distinguish between the
variety of models able to represent the average overall continuum in a typical
Seyfert galaxy (e.g. Loska \& Czerny 1997). 

However, the continuum models can be tested against the data only 
if we are able
to determine the observed continuum emission of the nucleus. It means that
the observed continuum has to be decomposed into true nuclear emission and
contributions from circumnuclear components.

This additional continuum emission clearly consists of the (variable) 
contribution from the Balmer continuum and blended FeII lines emitted by the
Broad Line Region (BLR) (e.g. Wills, Netzer \& Wills 1985) 
and from the stars located at a distance 
of a few 
hundreds of parsecs from the nucleus (e.g. Yee \& Oke 1978). 
It was suggested, however, that also
the third component is present, much bluer than the 
starlight and rather similar to the true nuclear continuum but coming from
extended region (e.g. Antonucci 1993, Tran 1995). 
Since both this component and the true
nuclear emission are featureless, the two components are called FC2 and FC1,
correspondingly. 

In this paper we attempt to make a decomposition of the optical continuum 
of NGC~5548 derived during the monitoring campaign
into nuclear emission and the remaining three components listed above. 
Whenever possible, we
use two methods at every stage,  
to ensure that the decomposition does not rely
strongly on the adopted method. 
Such a decomposition opens the way to subsequent
modeling of the variability of the nuclear continuum.

The outline of the paper is the following. In Section 2 we discuss the
contribution of starlight to the data and we derive its relative normalization
from aperture effects as well as from the depth of absorption features.
In Section 3 we derive the contamination by the FC2 component from the relation
between the  $H_{\beta}$ line and the true nuclear continuum at 5100~\AA.  
Finally, the true 
nuclear emission in the optical/UV band is presented in Section 4. We compare
it with the predictions of 
the model of an accretion disk with accreting corona. 
The discussion is given in Section 5.  

\bigskip
\centerline{\bf 2. Contamination by Starlight}
\bigskip

The optical data collected during the monitoring campaign are described
in a series of papers by Peterson et al. (1991,1992,1994). The data are
available to the astronomical community due to the courtesy of the AGN Watch team.
The spectra are calibrated using standard software for a corresponding
telescope. However, as different telescopes and apertures were used, 
secondary   calibration is needed in order to reduce all the spectra to the
set A with aperture $5.0" \times 7.6"$. 
We performed this reduction following the description given by
Peterson et al. (1994), i.e. we used values of $\phi$ 
(a point-source correction factor)
 and $G$ (starlight correction factor at 5100~\AA) appropriate 
for a given set of data,
calibrated to the mean [OIII] ($\lambda$5007) flux.  
We transformed the spectra to the 
rest frame and we also corrected the spectra 
for the effect of interstellar extinction in the Galaxy. The reddening is not
significant -- we adopted the value of E(B-V) equal 0.05, as inferred from 
the neutral hydrogen column
(Burstein \& Heiles 1984)
in the direction of this galaxy, although even smaller value 0.035 was 
recently
used by some authors (e.g. Reynolds 1997). 
We
deredden the spectra using the standard Seaton extinction curve (Seaton 1979).
It results in the increase of the flux at 5100~\AA ~ by a factor 1.174 and
at 1350~\AA ~ by a factor 1.499. We adopted these values in further
computations. 
Spectra prepared in such way were used to determine the amount of starlight 
in the standard set A aperture.  

\bigskip
{\it 2.1. Stellar Absorption Lines}
\bigskip

The fraction of light due to stars can be estimated from the ratio of
the observed equivalent width of stellar absorption lines to their average 
value in the spectrum of a template galaxy, which we take to be the spectrum
of the nuclear bulge of M31, after Coleman et al. (1980), Table 2. We have
used this galaxy as a template since it was shown by Wamsteker et al. 
(1990) that the starlight spectrum of NGC~5548 is similar to the 
starlight spectrum of M31. 
 
For the purpose of starlight estimation the spectrum n58733h from day
8733 (JD minus 2,440,000)
 from set H of Peterson et al. (1994)  was used. As  spectra from this set have
the broadest wavelength coverage (together they cover from 3000~\AA~
to 10,000~\AA),  more absorption features could be identified. 
It was also important that the spectrum  came from the epoch at which
the galaxy reached its minimum so the contribution of starlight 
to the total flux was more pronounced. At day 8733
the nucleus  was rather dim  with the 
total flux at 5100~\AA~ of $6.48 \times 10^{-15}ergs/s/cm^{2}/$\AA~.  
The  minimum of $F_{\lambda}$ during the entire 1988-1994 campaign 
was equal to $5.70 \times 10^{-15} ergs/s/cm^{2}/$\AA~ (day 8816).

The n58733h spectrum was normalized
by dividing the flux by the local value of continuum (obtained by fitting a 
spline to the spectrum). A similar procedure was applied to the template
starlight galaxy. This method removed broad band trends from the 
continuum of these spectra and showed the absorption features more clearly. 
In Figure 1 the normalized spectrum n58733h is presented together with  
the normalized template galaxy M31.

The spectroscopic indicators of stellar population are such prominent
absorption lines as:  
the CaII H ($\lambda$3968) and CaII K ($\lambda$3934) line and 
NaI D ($\lambda$5890,5896) line which are often
contaminated by interstellar absorption in the host galaxy. 
Other absorption features such as
 the CaII G ($\lambda$4304) and the MgI b triplet ($\lambda$5167,5173,5184) 
are usually contaminated by the
nearby emission lines (the CaII by  $H_{\gamma}$ and the MgI by
NI) so these lines were not used for starlight estimation. 
The lines that were used 
are: FeI+CaI ($\lambda$6129,6072)  and FeI ($\lambda$7406,7458,7511). 
In Table 1 we show the percentage of starlight estimated from different 
absorption features  and defined as the EW of an absorption 
feature in the spectrum to the EW of the same feature in the 
template starlight galaxy M31. Next we calculate the 
amount of starlight at 5100~\AA ~ in absolute units taking into account the
spectral shape of both the starlight and the combined starlight plus nuclear 
emission. The value 
estimated from the FeI+CaI ($\sim$6100~\AA) feature is 
$3.4\times 10^{-15}ergs/s/cm^{2}/$\AA~ and estimated from the FeI
($\sim$ 7500\AA) feature is $3.3\times 10^{-15}ergs/s/cm^{2}/$\AA.

The estimate of starlight contribution based on spectral features does not
include the possible contribution from featureless FC2 component. It is 
therefore particularly interesting to compare it with the determination of
starlight by aperture methods which may include most of its contribution if
this component is extended. 

\bigskip
{\it 2.2. Comparison with Aperture Estimates}
\bigskip

Another estimation of starlight  can be made by comparing
spectra obtained through different apertures. By subtracting 
a  small aperture spectrum from a large aperture spectrum we can
obtain off-nuclear emission from the circumnuclear ring which 
should be entirely due to starlight. We found a few pairs of   
observations made with distinctively different apertures and 
separated in time not more than four days, so the  variations of the flux 
were relatively unimportant. It was also crucial for the different
aperture spectra to have  large wavelength coverage as to 
obtain the shape of starlight over a broad wavelength range. 
Spectra which best fitted the above criteria were 
those from set F (aperture $3.2" \times 6.4"$) 
and H (aperture $4.0'' \times 10.0$'') which together covered
4500~\AA~ to 7000~\AA. An example of an off-nuclear spectrum obtained 
from the subtraction of spectra from these sets (n58765h minus
n58765f) is shown in Fig. 2, together with the template galaxy M31 used in the 
Section 2.1.  The
"starlight" obtained in such a way is 
much steeper than the template starlight and shows a prominent $H_{\alpha}$
line. The cause  of such  steepening is the fact  
that in the Peterson et al. (1994) data the spectra at different apertures 
were obtained with different telescopes and most  probably  
the original reduction of the data from the small aperture  
(the 1.6~m Mount Hopkins observations) was much too simplistic    
(D. Dobrzycka, private communication).  

The available data did not allow for the determination  of  starlight 
basing on different aperture spectra.  
It also showed that broad band optical spectra are not calibrated well
enough to use for comparison with the models. In further Sections we constrain
ourselves to spectra from  set A or fluxes at 5100~\AA ~ for the 
purpose of modeling. Because data from these sets do not cover the
wavelength range of the Balmer continuum and the FeII emission lines
we do not need to worry about this contribution in the spectra in 
further analysis.

Much better attempt was made by Romanishin et al. (1995), who using a single 
telescope  made direct imaging of NGC 5548. The authors obtained the
value of $3.4\times 10^{-15} ergs/s/cm^{2}/$\AA~ for
starlight at $F_{\lambda}$(5100\AA)  (for aperture 5"$\times$7.6") which
is identical with the value of $3.37\pm0.54\times10^{-15} 
ergs/s/cm^{2}/$\AA~
found from comparison of IUE and 5100~\AA~ fluxes by Peterson 
et al. (1991). This method gave the same value of 
starlight as the method based on stellar absorption
features (see previous section).

The method of estimating starlight based on absorption features
is more reliable in the case of using high quality 
nuclear and off-nuclear spectra of the same galaxy obtained
from the same telescope (e.g. Serote-Ross et al. 1996 for NGC~3516).
For NGC 5548 off-nuclear data are not available and simple use of templates
from another galaxy may lead to highly uncertain results.

However, since the outcome of our absorption feature method coincide well 
with the aperture
based results of Romanishin et al. (1995)  we can safely use the value 
$3.4 \times 10^{-15} ergs/s/cm^{2}/$\AA~ 
of starlight in further analysis.

\bigskip
\centerline{\bf 3. Contribution from Extended Featureless Continuum FC2}
\bigskip

The monitoring of NGC 5548 showed that the continuum of 
this galaxy as well as the $H_{\beta}$ flux
varies strongly with time. We use this phenomenon to estimate the amount
of the extended featureless continuum FC2 within the standard set A spectra. 

The broad $H_{\beta}$ line is found to vary in response
to continuum variations with a lag of about 20 days, which varies from year
to year.  Figure 3
presents the dependence of $H_{\beta}$ flux (corrected for Galactic reddening) 
from optical flux at 5100 \AA 
(corrected for Galactic reddening and starlight contamination). The 
$H_{\beta}$ 
fluxes were shifted in time by the mean delay of the campaign equal 18 days.

Apart from the short sequence marked with large circles (dates from 
7728 to 7766), the line responds almost linearly to the underlying continuum.

The $H_{\beta}$ line consists of components originating in the broad line
region and in the narrow line region. The 
component that is sensitive to short-time nuclear continuum changes is the 
component
that lies nearest to the central engine, i.e. the $H_{\beta}^{BLR}$. The 
remaining $H_{\beta}$ components originate from the Narrow Line Region 
(NLR) that lies further away
(the distance to the NLR is of order of a hundred light years) 
and is not subject
to such changes. 

We estimated the value of $H_{\beta}^{NLR}$ by decomposing the line profile
into two components: a broad line and a narrow line (see Figure 4). 
We used for that purpose the spectrum n58733a (day 8733) 
as the broad components  during that epoch were  weak and the 
relative contribution from the narrow components was  high, 
which allowed for relatively accurate decomposition. The value of 
$1.5 \times 10^{-13} erg/s/cm^2$ for the narrow component was obtained. 

We compared this value with the value expected on statistical grounds. 
The amount of $H_{\beta}$ originating from the narrow line region 
can be estimated based on paper by Osterbrock (1989), who showed 
that the logarithm of the ratio of $H_{\beta}$ to [OIII] flux in the 
NLR for Seyfert galaxies is in the  -0.3 to 1 range. 
The value of $0.35-11.1\times10^{-13}ergs/s/cm^2$ 
 is obtained assuming that the 
[OIII] flux for NGC 5548, as inferred from 
large-aperture observations (Peterson et al. 1991), is $5.58\times10^{-13}
erg/s/cm^{2}$. 
Since the observed angular size corresponds to 340 pc per 
arcsec (z=0.0174, d=70 Mpc for $H_{o}$=75 km/s/Mpc) 
the size of the region observed in the standard A aperture is
1.7kpc$\times$2.6kpc  which covers the whole narrow
line region.  Therefore  the value of $H_{\beta}^{NLR}$ derived from the
decomposition of the line originates in the NLR and  is well within 
the expected limits.

We expect that the line should change in agreement with variations of the
nuclear continuum $F_{nucl}$ as follows: 
	$$H_{\beta}(t)=H_{\beta}^{NLR} + A F_{nucl}(t) \eqno(1)$$
However, as the observed flux consists of the variable nuclear  part and 
constant contribution of the FC2 component, the observed relation is
the following:
$$H_{\beta}(t)= A (F_{nucl}(t) + F_{FC2}) + B. \eqno(2)$$
Comparing the two formulae we can estimate the contribution from FC2 component
$$F_{FC2} = (H_{\beta}^{NLR}-B)/A. \eqno(3)$$

The values of the coefficients $A$ and $B$ are derived directly from the fit
of the straight line to the data (Fig.3).
The value of the coefficient B  is 
$1.95 \times
10^{-13} erg/s/cm^{-2}$ and A is equal 0.92 in appropriate units.

The value of $H_{\beta}^{NLR}$ equal 1.5$\times10^{-13}ergs/s/cm^{2}$  
gives us the contribution of the
FC2 component to the continuum at a level of 
0.45~$\times10^{-13}erg/s/cm^{2}$. It is
about 20 \% when the source is weak and drops down to a few per cent when
it is bright.  These values are in agreement with the estimate made by
Tran (1995), who showed that the FC2 continuum in Seyfert~
1 galaxies account for only about 4 \% of the total featureless continuum
(defined as the sum of nuclear continuum and the
scattered, polarized "featureless continuum") while in the Seyfert~2
 for 60\%-90\%. Therefore, even at its
low level, NGC 5548 is not truly approaching Seyfert 2 galaxy properties,
although the contribution from FC2 is not completely negligible either.

Unfortunately, we cannot say anything about the shape of the FC2 component
since the available off-nucleus spectra are not reliable (see Section 2.2).
Therefore, we have to ignore its contribution to the optical/UV spectrum
in the next Section.

\bigskip
\centerline{\bf 4. Comparison of the Intrinsic Nuclear Emission
with the Accreting Corona Model}
\bigskip

We use the determination of the intrinsic nuclear emission to
test a specific model of an active nucleus. That model consists
of an accretion disk surrounded by a corona, with both an
accretion disk and a corona powered directly by accretion (\. Zycki,
Collin-Souffrin \& Czerny 1995, Witt, Czerny \& \. Zycki 1997). 
It describes the
spectrum of the nuclear emission in the entire optical/UV/X-ray band. The
structure of the flow is calculated self-consistently at every radius, as
described by Czerny, Witt \& \. Zycki (1997), and the emerging
spectrum is integrated over the disk surface so the stationary model is
characterized by the global parameters: mass of the central black hole, 
the accretion rate and the viscosity parameter $\alpha$.

\bigskip
{\it 4.1. High and Low State Models}
\bigskip
            
We start with detailed modeling of the two spectra representing a
high luminosity state and a low luminosity state. We choose spectra from
set A since these are the most reliable (see Sect. 2.2).  In Fig. 5 
we show examples of
a low (n58733a) and a high (n57649a) flux spectra. The spectra are 
corrected for starlight, reddening and transformed to rest frame. 
Unfortunately, they cover only a relatively narrow energy 
range. The IUE data are available for high state only. In order to show  
an example of the spectrum covering  the entire optical band but to avoid the
problem of starlight subtraction in these not well calibrated data we
 plot (see Fig. 6) the difference 
between the high and low states from set H (n58733h subtracted from n57643h). 

The difference between the two states is considerable.
For higher states the blue bump becomes more prominent, the $H_{\beta}$
line broader and the underlying continuum changes slope. 
The data for high and low states are compared with a model of an 
accretion disk with a hot accreting 
corona. We adopt the value of the viscosity parameter $\alpha$ equal 0.1,
as suggested by statistical data for quasars (Czerny, Witt \& \. Zycki 
1997). For such a
 low viscosity the radial infall of mater is accompanied by
strong vertical outflow which was included in the model, as described by 
Witt et al. (1997). The best fit to the observed data was obtained for the 
following
parameters: mass of the central black hole
$M_{BH}=1.1\times10^{8}M_{\odot}$,
and the following accretion rates:
\.{m}=0.05 for the high state and  \.{m}=0.03367 for the low state (Fig.5).
The difference between the high and low states resulting from our 
computations is also plotted against the observed difference (Fig. 6).

In the case of high state the effect of mass loss does not have an important
influence on the spectrum. However, in the case of low state the effect is
profound so the external accretion rate is only slightly lower than in the
high state but the differences in the bolometric luminosities and spectral
shapes  between the two states are considerable. 

The model gives a good fit to the observational data. It 
also predicts the X-ray flux in the 2-10 keV region.
The estimated value of the 2-10 keV flux obtained from the model 
for the high state
is: $2.04\times10^{-11} ergs/s/cm^{2}$. 

This estimate can be compared with a prediction based on observational data 
from Clavel et al. (1992). The observed (undereddened) UV flux at
$\lambda=1350$\AA~  (log$\nu=15.35)$ at that epoch was  
equal to 5.82$\times10^{-14} ergs/s/cm^{2}/$\AA.
For $ F_{\lambda}$(1350\AA) of the above value the linear formula gives 
the X-ray flux equal  6.64$\times10^{-11}ergs/s/cm^{2}$. 
 
The model therefore is fainter in X-rays than it is expected. Although the
linear formula clearly overpredicts the expected value of X-ray flux, 
as there is a strong flattening in the dependence of the X-ray brightness 
from the UV flux towards higher UV luminosities, still the required value
should be about 5$\times10^{-11}ergs/s/cm^{2}$, as suggested by visual 
inspection of Fig. 4 from Clavel et al. (1992).

Higher X-ray luminosities for given UV flux are expected from the model if the
viscosity parameter is smaller. Adopting $\alpha = 0.05$ we can fit the 
high state data with the parameters: $M_{BH} = 1.4 \times 10^8 M_{\odot}$, 
$\dot M=0.037 M_{\odot}$/yr and predict the 2-10 keV flux equal 
3.65$\times10^{-11}erg/s/cm^{2}$, closer the expectations based on observed
trends. However, in that case the required value of the mass of the central
black hole becomes somewhat larger than the values between 
$7 \times 10^7 M_{\odot}$ --  $10^8 M_{\odot}$ which resulted from estimates 
based on the study of the Broad Line Region 
dynamics (Wanders et al. 1995, Done \& Krolik 1996; see also
Loska \& Czerny 1997).

\bigskip
{\it 4.2. Predicted Dependencies on Accretion Rate}
\bigskip

In this Section we compare the overall time behavior of the nucleus with
predictions of the accreting corona model. We fix the value of $M$ of 
the mass of the
central black hole and calculate a sequence of 
stationary models parametrized by the accretion rate with the spectra 
covering optical/UV/X-ray band. Therefore, we obtain a 
theoretical relation between 
the  fluxes  at  5100~\AA,  1350~\AA~ and the integrated
2 - 10 keV luminosity, for each $M$.
We compare it with the
data from the first year of the observational campaign, when the IUE data
were available and with the data of Clavel et al. (1992).

The value of the mass
$1.1 \times 10^8 M_{\odot}$ suggested by the fit to the optical data in a
single high state data set does not reproduce well the overall trend 
(see Fig. 7). Therefore, we also plot a theoretical line corresponding to
the central mass of $8 \times 10^7 M_{\odot}$ and the viscosity parameter 
equal 0.1, as before.

We see that the  main trend might be well reproduced assuming a value 
of the mass of the central black 
hole about $9.5 \times 10^7 M_{\odot}$ which is 
actually closer to the value obtained from
dynamical considerations. However, smaller mass models lead to even
stronger underproduction of the X-ray brightness than higher mass 
models (see Section 4.1). 

To inspect the problem of underproduction of X-rays we construct a theoretical
diagram of the dependence between the UV flux and the integrated flux between
2 and 10~keV (Fig. 8). We add the observational points from Clavel et al. 
(1992). Both families of models from Fig. 7 underproduce the X-ray emission 
and the predicted character of the dependence is in contrast with 
the observational points.

Therefore we modified our theoretical model. We neglected the effect
of mass loss predicted by the model and used the same value of the accretion 
rate at all radii for a given external accretion rate. This led to a 
qualitative change of the shape of the dependence between UV and X-ray flux.
We adjusted the mass of the central black hole to represent well the
relation between the optical and UV flux (dashed line in Fig. 7).
Additionally, by taking smaller value of the viscosity parameter $\alpha$ we
could adjust the level of the X-ray emission to the required level (see 
dashed line in Fig. 8).

Our modified model of the accreting disk with accreting corona is now 
consistent with the behavior of all three spectral bands: optical, UV and 
X-ray, remaining still a two-parameter model, described by the central mass
and the viscosity. In the case of NGC 5548 these parameters are: $M_{BH}=1.4
\times 10^8 M_{\odot}$ and $\alpha = 0.033$. 

The spectral range in the optical data without calibration problems (set A)
is too short to attempt the distinction between the models with and without
mass loss. The model without mass loss, $M_{BH}=1.4
\times 10^8 M_{\odot}$, $\alpha = 0.033$. and accretion 
rate $\dot m=0.009$ cannot be visually 
distinguished from the model given in Fig. 5 (i.e. with mass loss, 
$M_{BH}=1.1
\times 10^8 M_{\odot}$, $\alpha = 0.1$. and accretion 
rate $\dot m \approx 0.034$).

We see that the short sequence of points marked with large circles (dates from 
7728 to 7766) in Fig. 7 follows approximately the overall trend of 
the continuum variations 
although the $H_{\beta}$ line response was untypical. Nevertheless, simply 
on the basis of Fig. 7 we cannot tell what caused the departure of those
points from the main trend shown in Fig. 3. Two possibilities are open. 
Either the line flux
was too high (i.e. the BLR did not respond to the decrease of the nuclear
emission although the considered period was longer than the usual time
delay), or the continuum flux seen by us was too low. This second possibility
would happen if some intervening optically thick matter crossed our line of
sight.  

Although the model of accreting corona is able to reproduce well the broad 
band  time behavior of the nucleus,
two problems remain. The first is the question, why the mass loss predicted by
the model is possibly too large. The second is that the favored value of the
central mass is somewhat larger than suggested by the dynamical studies of the
Broad Line Region, although the discrepancy is not stringent.

\bigskip
\centerline{\bf 5. Discussion and Conclusions}
\bigskip

\bigskip
{\it 5.1. Data Requirements}
\bigskip

The character of variability of the Seyfert galaxy NGC 5548 is very complex. It
means that the observational data not only have to be dense in time, in order
to cover well the short time scale behavior but  also have to extend for
long times (years) to catch the important long time scale trends. This 
general problem shows up in a number of particular difficulties.

A few observational points do not always follow the general trend. Two 
observations made when the emission lines in NGC 5548 were exceptionally weak
suggested that the almost complete disappearance of the lines is a break off
from the usual behavior of the source (Iijima 1992, Iijima, Rafanelli \&
Bianchini 1992, Loska, Czerny \& Szczerba 1993).  Actually, the behavior of
the source during that period followed well the general trend, as shown by
the continuous frequent sampling (Peterson et al. 1994; see also Figure 3).

On the other hand, there seem to be periods when the source does not follow
the typical variability pattern. During the period of about six weeks
in July/August 1989 (days 7722 to 7766) the continuum was much fainter 
than expected on the
basis of the $H_{\beta}$ intensity. Since the line delay (about 18 days) 
is much shorter than six weeks the effect cannot be caused by the lack of
answer of the Broad Line Region to the change of the continuum. It may
suggest some kind of geometrical effect which decreased the level of 
continuum seen by an observer but not by BLR. However, it may also indicate
that the coherent response of the BLR require more time than indicated by
the measured time delay. In order to  avoid problems,  
we excluded the data collected during that period from our modeling. 

In our comparison of the data with the model the observational point in X-rays
obtained in 1985 plays a crucial role since the amplitude of the variation 
reached  exceptionally large value at that time. Such events are rare in a 
sense that large variations happen in time scales of years but their
inclusion in the study is crucial from theoretical point of view so the
good data have to cover such long time scales as well as short time scale
variability.

The presence of the broad and continuous range of time scales in the problem is
best seen in the study of X-ray power spectra which are basically of a 
power law shape in all AGN (McHardy \& Czerny 1987, Lawrence et al. 1987;
see also Czerny \& Lehto 1997). It is not clear whether the long term
variability (in the time scale of years) is actually of the same nature that
the best studied variability in time scales of hundreds to thousands of
seconds but we cannot a priori reject such a possibility and the large
amplitude variations like the one mentioned above should be included in
any model unless there are arguments against it.

The currently available data on NGC 5548 are already good enough to allow
interesting study of the time dependent properties but extension of the
observational campaign is essential for better understanding of the behavior
of accretion flow.

\bigskip
{\it 5.2. Determination of Intrinsic Spectrum of NGC 5548}
\bigskip

The results of Section 2 show that the level of starlight at 5100 \AA ~
(in the aperture 5.0'' x 7.6'') equal to 
$3.4 \times 10^{-14}$ ergs/s/cm$^2$/\AA~ 
is determined reliably. However, the spectral shape of the starlight does
not seem to be described accurately. Separate spectrophotometric 
observations with a good telescope are necessary in order to solve 
this problem.

The lack of good starlight determination in the entire optical band limits
the testing of the model against the data basically to the normalization
at 5100 \AA~. It is not a big problem if the entire optical/UV/X-ray range
is studied but nevertheless it makes impossible to use the entire optical range
for fine tunning. 

The subtraction of the FC2 component is another problem which should be dealt
with while fine tunning the model since the level of the FC2 at 5100 \AA~
is up to 20\% when the source is exceptionally faint. Fortunately, the source
is brighter most of the time so the FC2 contribution could have been neglected
in the present analysis of the applicability of the model.

\bigskip
{\it 5.3. Model Advantages and Restrictions}
\bigskip

The variations of the intrinsic continuum emitted by the nucleus of NGC 5548
in the optical/UV/X-ray band were successfully reproduced using a simple
model of an accretion disk with an accreting corona by fixing the
mass of the central black hole at $1.4 \times 10^8 M_{\odot}$, the viscosity
parameter at 0.033 while varying accretion rate from $\sim 0.009 $ to
$\sim 0.04 M_{\odot}$/yr.

The model reproduces well both -- the hardening of the spectrum
with an increase of the luminosity and the weak increase in X-ray luminosity
with the increasing UV flux when the source is bright. The disk emission 
does not extend into soft X-rays so any traces of the soft X-ray excess
in this source (Walter \& Fink 1993) have to be attributed to the atomic
processes connected with reprocessing of X-rays by the disk surface (Czerny
\& \. Zycki 1994). We neglected those processes in our model assuming that 
the disk albedo is equal to zero.

The version of the model which corresponds well to the behavior of the 
source is the model in which the vertical outflow of the mass from the
corona in the form of the wind is neglected. The theory underlying the
description of the outflow will have to be reconsidered. Apart from that,
the model is a very promising one and should be further tested and developed.

The most important topic is the understanding of the nature of the variations
and the character of the global response of the disk/corona system. 

In our
present approach we simply use a set of stationary models characterized by
a single accretion rate at all disk radii at any given moment. This is
certainly an oversimplification. The variations in the optical/UV band are
noticeable starting from the time scales of one -- two days (Korista et al. 
1995). The main contribution to the spectrum comes from the inner parts of
the disk, at radii $r \sim 10~R_{Schw}$. The thermal time scale of that
region depends on the viscosity parameter $\alpha$ and  is equal
$$\tau_{th} = 0.2 ({ 0.033 \over \alpha})({ M \over 1.4 \times 
10^8 M_{\odot}}) ~~~~ {\rm days}$$
and a more accurate value of the two-folding time scale at 1350 \AA~ 
(independent from the assumption about the disk
radius dominating the emission at a given frequency) is equal 1.0$^d$ for the
mean luminosity level (Siemiginowska \& Czerny 1989). Therefore, if there are
any changes of the corona structure the thermal structure of the disk can
follow these changes over the time scales of days. 

The viscous time scale is longer than the thermal time scale by a few orders of
magnitude (see e.g. Siemiginowska, Czerny \& Kostyunin 1996). The disk 
therefore is never stationary and our assumption is not justified. On the
other hand this assumption is the only reasonable approach since we cannot
calculate the actual time behavior of the disk because of the lack of 
proper understanding of the disk/corona coupling. The disk/corona system 
described by the model is thermally unstable (see Fig. 7 of Witt et al. 1997).
The sudden increase of the corona activity can therefore result even in 
enhanced disk response, as required, but this response does not have to
saturate at the value of the flux radiated by the disk which is necessary
for the corona to reach the new temporary thermal balance. However,
if we do assume that
the thermal disk runaway saturates at the appropriate emitted flux then the
disk/corona structure would be observationally the same as in the case
of enhanced accretion rate since the disk/corona coupling does not depend
on cold disk structure (surface density etc.).

The applicability
of the stationary model to the description of the  time evolution suggests that
such a saturation actually takes place but its mechanism is unknown. The
additional disk/corona coupling 
might be based on magnetic field and the description of the disk and corona
flow by the viscosity coefficient $\alpha$ is too simple to reproduce well the
time behavior, including stability,  
although may approximate the quasi-stationary 
situation. 

\bigskip

ACKNOWLEDGEMENTS. This research was based on the optical data collected
by the AGN Watch team available through private communication with
B.M. Peterson. We are grateful to Danusia Dobrzycka, Joanna Miko\l ajewska
and Wojtek Dziembowski for helpful discussions. 
The project was partially supported by 
the Polish State 
Committee for Scientific Research grant no. 2P03D00410.

\bigskip
\centerline{REFERENCES}
\bigskip

\Referencja{Antonucci, R., 1993, Ann. Rev. Astron. Astrophys., 31, 473}
\Referencja{Burstein, D., Heiles, C., 1984, ApJS, 54, 33}
\Referencja{Clavel, J. et al., 1991, ApJ, 366, 64}
\Referencja{Clavel, J. et al. 1992, ApJ, 393, 113}
\Referencja{Coleman, G.D., Wu, C-C., Weedman, D.W., 1980, ApJ. Suppl. 43, 393}
\Referencja{Czerny, B., Witt, H.J., \. Zycki, P.T., 1997, in  
   {\it ESA SP-382}, 2nd INTEGRAL Workshop,``The Transparent Universe'', 
    eds. C. Winkler, T. Courvoisier and Ph. Durouchoux, page 397}
\Referencja{Czerny, B., Lehto, H.J., 1997, MNRAS, 285, 365}
\Referencja{Czerny, B., \. Zycki, P.T., 1994, ApJL, 431, L5}
\Referencja{Done, C., Krolik, J.H., 1996, ApJ, 463, 144}
\Referencja{Done, C., Pounds, K.A., Nandra, K., Fabian, A.C., 1995, MNRAS, 
     275, 417} 
\Referencja{Edelson, R., et al. 1995, ApJ, 438, 120}
\Referencja{Iijima, T., 1992, IAU Circ. 5521}
\Referencja{Iijima, T., Rafanelli, P., Bianchini, A., 1992, A\&A, 265, L25}
\Referencja{Korista, K.T., et al, 1995, ApJS, 97, 285}
\Referencja{Lawrence, A., Pounds, K.A., Watson, M.G., Elvis, M.S., 1987,
       Nat., 325, 692}
\Referencja{Loska, Z., Czerny, B., Szczerba, R., 1993, MNRAS, 262, L31}
\Referencja{Loska, Z., Czerny, B., 1997, MNRAS, 284, 946}
\Referencja{Marshall, H.L., Fruscione, A., Carone, T.E., 1995, ApJ, 439, 90}
\Referencja{McHardy, I.M., Czerny, B., 1987, Nat, 325, 696}
\Referencja{Osterbrock, D.E., 1989, in Astrophysics of Gaseous Nebulae and
Active Galactic Nuclei p. 346}
\Referencja{Peterson, B.M., et al. 1991, ApJ, 368, 119}
\Referencja{Peterson, B.M. et al. 1992, ApJ, 392, 470}
\Referencja{Peterson, B.M., et al. 1994, ApJ, 425, 622}
\Referencja{Reichert, G.A. et al. 1994, ApJ, 425, 582}
\Referencja{Reynolds, C.S., 1997, MNRAS, 286, 513}
\Referencja{Romanishin, W., et al. 1995, ApJ, 455, 516 }
\Referencja{Seaton, M.J., 1979, MNRAS 187, 73P.}
\Referencja{Serote-Ross, M., Boisson, C., Joly, M., Ward, M.J., 1996, MNRAS,
            278, 897}
\Referencja{Siemiginowska, A., Czerny, B., 1989, MNRAS, 239, 289}
\Referencja{Siemiginowska, A., Czerny, B., Kostyunin, V., 1996, ApJ, 458, 491}
\Referencja{Stripe, G.M.. et al. 1994, ApJ, 425,609}
\Referencja{Tran, H.D., 1995, ApJ, 440, 597} 
\Referencja{Walter, R., Fink, H.H., 1993, A\&A, 274, 105}
\Referencja{Wamsteker, W. et al. 1990, ApJ, 354, 446}
\Referencja{Wanders, I., et al., 1995, ApJ, 453, L87}
\Referencja{Wills, B. J., Netzer H., Wills, D., 1985, ApJ, 288, 94}
\Referencja{Witt, H.J., Czerny, B., \.{Z}ycki, P.T. 1997, MNRAS,
           286, 848}
\Referencja{Yee, H.K., Oke, J.B., 1978, ApJ, 226, 763}
\Referencja{Zheng, W. et al. 1995,ApJ, 44, 632}
\Referencja{\. Zycki, P.T., Collin-Souffrin, S., Czerny, B., 1995, MNRAS,
          277, 70}

\bigskip\bigskip

\centerline {FIGURE CAPTIONS}
\bigskip

Fig. 1. The spectrum of NGC 5548 (n58733) (thin line) and the 
template starlight 
galaxy M31 (thick line), both normalized as described in Section 2.1.

Fig. 2. The off-nuclear spectrum of NGC 5548 calculated by subtraction of
the observation n58765f (aperture 3.2"$\times$6.4) from n58765h 
(aperture 4.0"$\times$10.0"). It differs considerably 
from the overall shape of the starlight template (given with arbitrary 
normalization).

Fig. 3. The dependence of the $H_{\beta}$ flux measured in units of
$10^{-13} $ erg/s/cm$^2$ on the continuum 
flux at 5100 \AA~  measured in units of $10^{-15}$ erg/s/cm$^2$/\AA ~
(after starlight subtraction). Both fluxes were dereddened. 
The data  cover the first four years of 
the campaign (Peterson et al. 1994). Points marked by large open circles 
come from the
four week sequence in July/August 1989 and were not included in the linear
fit to the trend (dashed line).

Fig. 4. Decomposition of the  $H_{\beta}$ profile into broad and narrow 
components in the data 8733, taken when the source was faint. A power law
continuum was adopted, normalised outside the spectral band dominated 
by strong emission lines. 

Fig. 5. A low (n58733a) and a  high (n57649a) states are shown, starlight
subtracted and dereddened and the IUE data points corresponding to the high
state. The fitted models are for 
$M_{BH}=1.1\times10^{8}M_{\odot}, \alpha=0.1$ and 
\.{m}=0.05, \.{m}=0.03367 for the high and  low state spectra,
respectively. Outer disk radius was put at 120 $R_{Schw}$.  Fluxes are given
in erg/s/cm$^2$/Hz.

Fig. 6. A broad band difference between two states (n58733h subtracted
from n57643h) is shown (in order to avoid stellar subtraction problem), 
together with the difference between two 
theoretical spectra from Fig. 5.  

Fig. 7. The relation  between  the optical and UV flux during the first year
of observational campaign given by Clavel et al. (1991) (squares) and during 
the fifth year given by Korista et al. (1995) (triangles),
after dereddening of the data. Points from July/August 1989 period with 
exceptionally high $H_{\beta}$ are marked by  large circles. 
Continuous lines show the dependence predicted by 
the model assuming the viscosity parameter equal to 0.1, for  two masses of the
central black hole. The dotted line marks the predictions for mass loss 
excluded, black hole mass equal to $1.4 \times 10^8 M_{\odot}$ and  
viscosity parameter equal to 0.033. Optical flux 
is given in units of $10^{-15}$
erg/s/cm$^2$/\AA, and UV flux in units of $10^{-14}$
erg/s/cm$^2$/\AA.

Fig. 8. The relation between the UV flux and the 
integrated 2-10 keV flux.
Data points are from Clavel et al. (1992), without dereddening. The brightest
point represents the measurement made in 1985. 
Lines mark the same sets of models 
as in Fig. 7, but reddened by Galactic extinction. 
UV flux is given in units of  $10^{-14}$
erg/s/cm$^2$/\AA~ and X-ray luminosity in units of $10^{-11} $ erg/s/cm$^2$/Hz.

\bigskip\bigskip

TABLE 1
$${\vbox
{\settabs 3 \columns
\+ starlight feature & ~~~~~~~~~~\% of starlight & remarks\cr
\hrule
\+ \cr
\+ FeI(4923, 5018,5169), MgI+MgH &~~~~~~~~~~ 5.8 \% & MgI contaminated by NI\cr
\+NaI(5890,5896) & & emission\cr
\+FeI+CaI(6072,6129) & ~~~~~~~~~~52.8 \% \cr
\+CaH(6934),FeI(7008,7109,7157) & ~~~~~~~~~~12.9 \% & CaH contamination by ISM\cr
\+FeI(7459,7511) & ~~~~~~~~~~64 \% \cr
\+ CaII (8498,8542) & ~~~~~~~~~~11 \% & contamination by ISM\cr
}}$$



\epsfbox{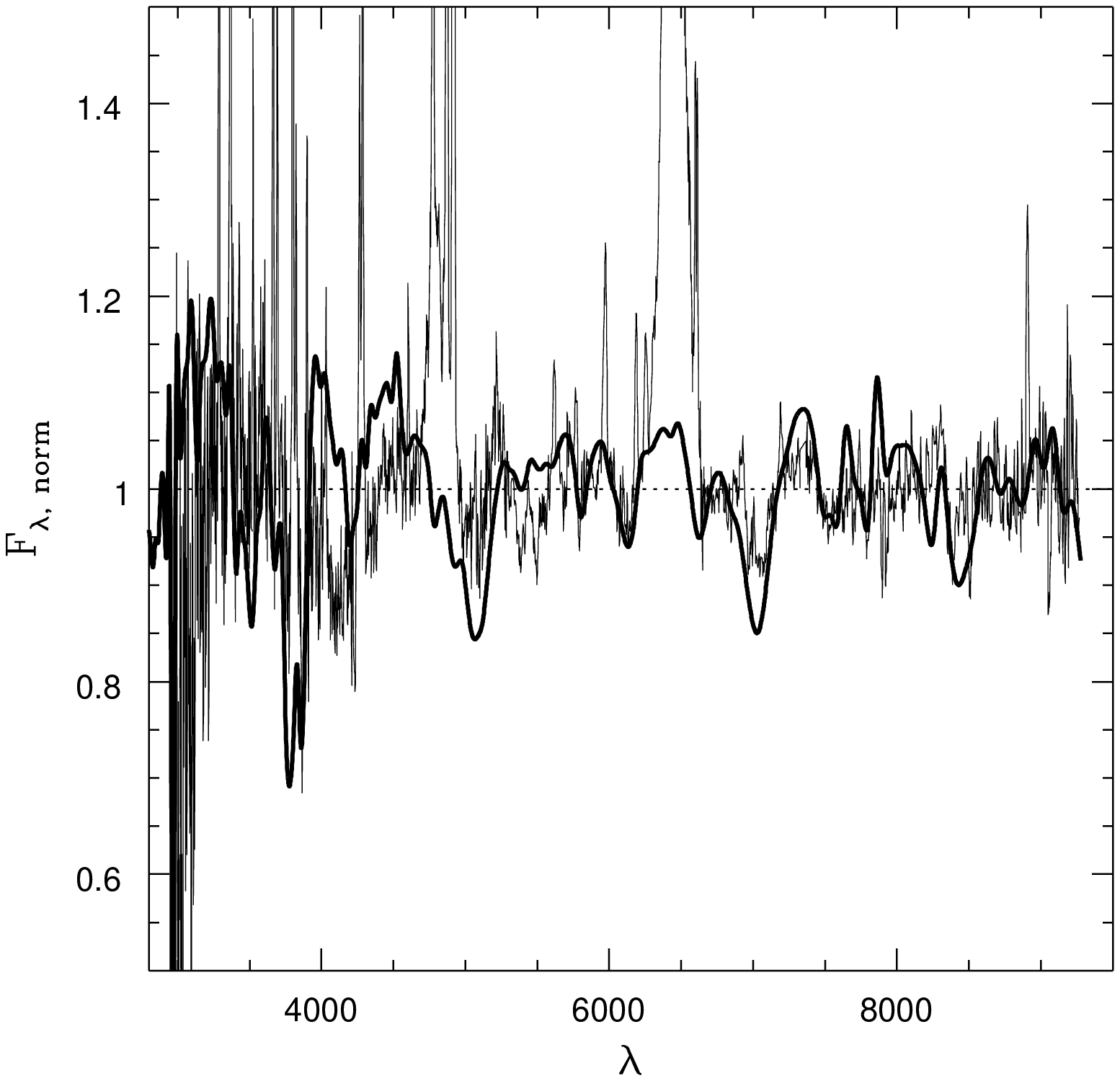}


\epsfbox{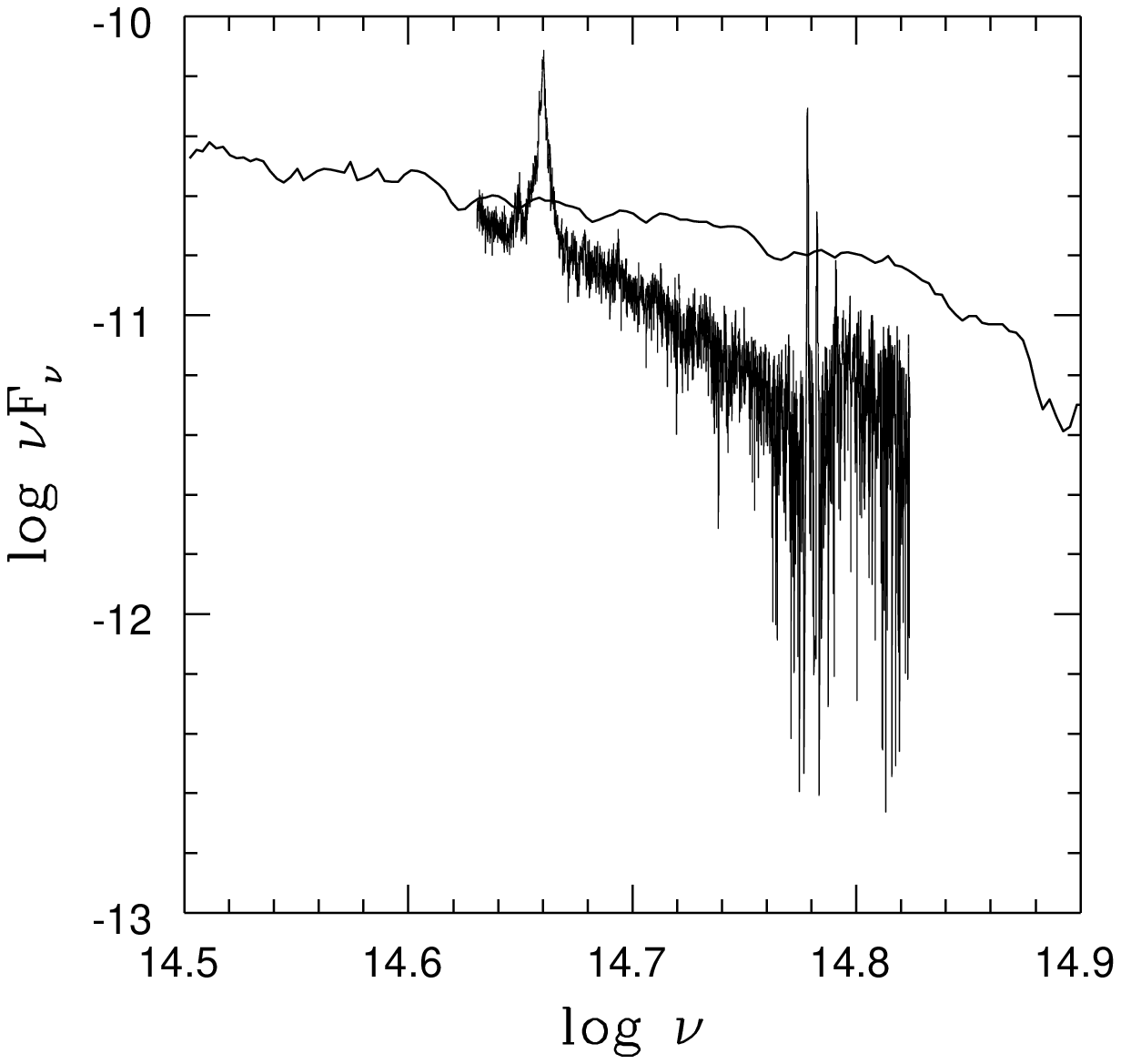}


\epsfbox{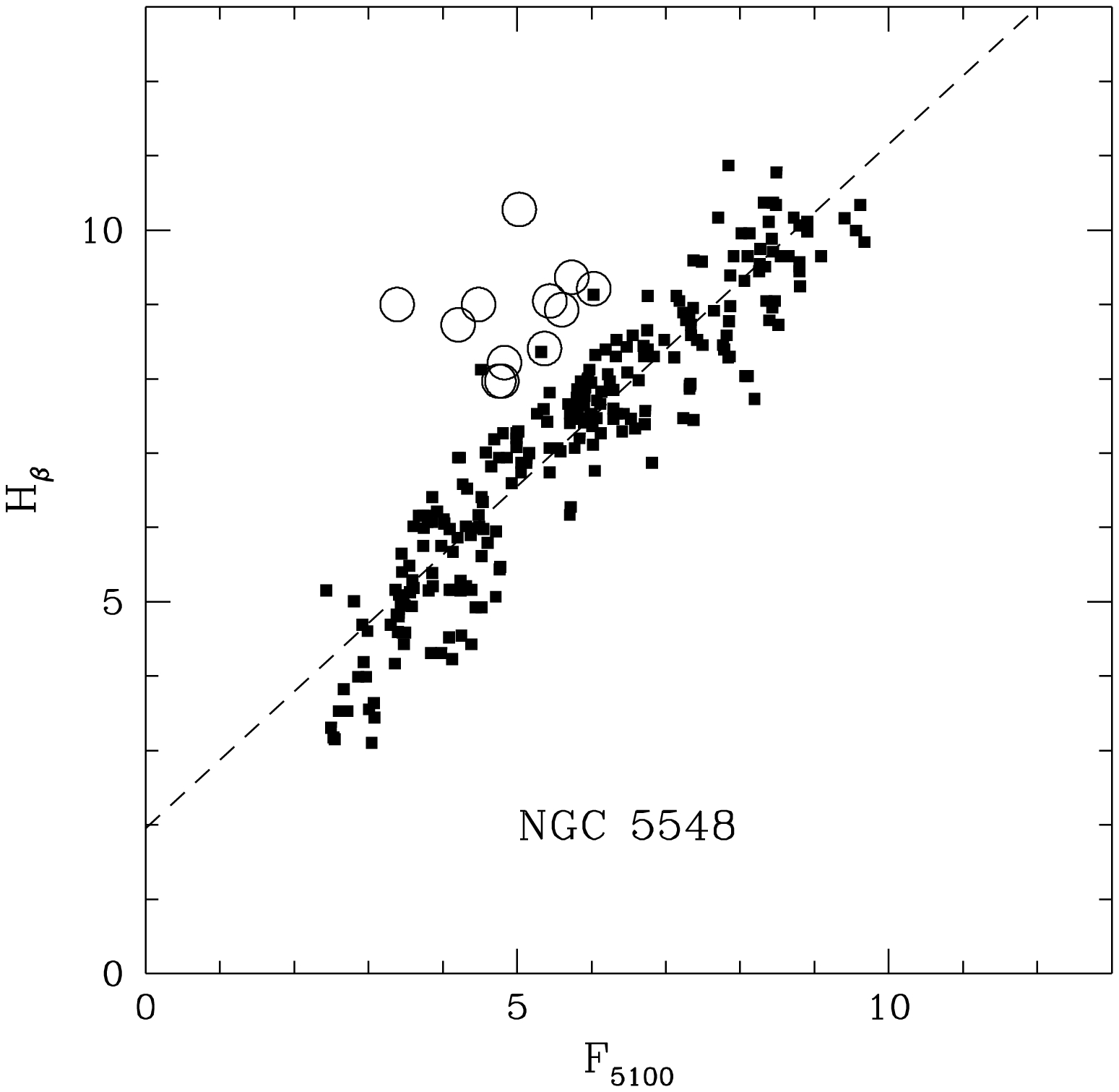}


\epsfbox{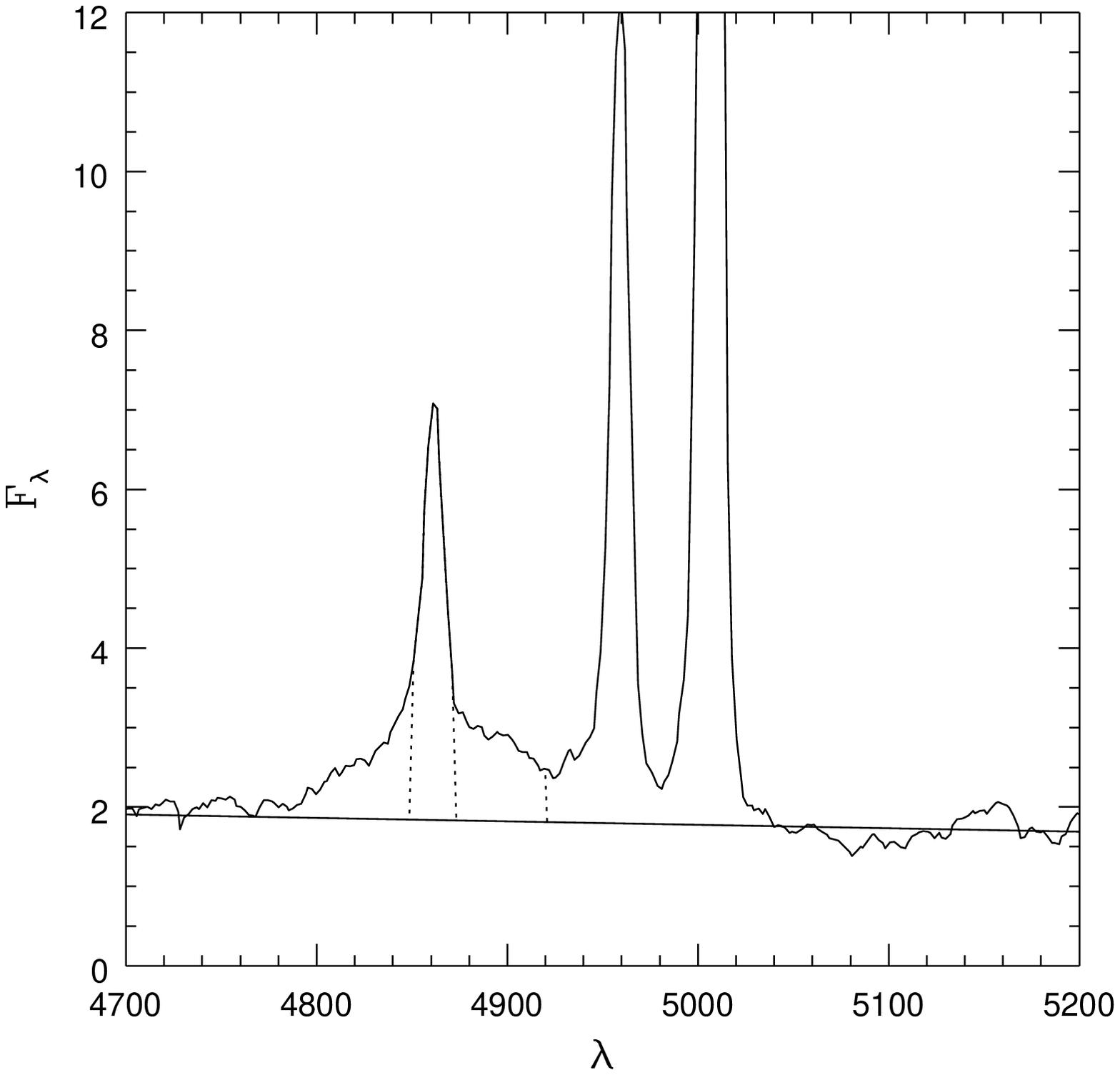}


\epsfbox{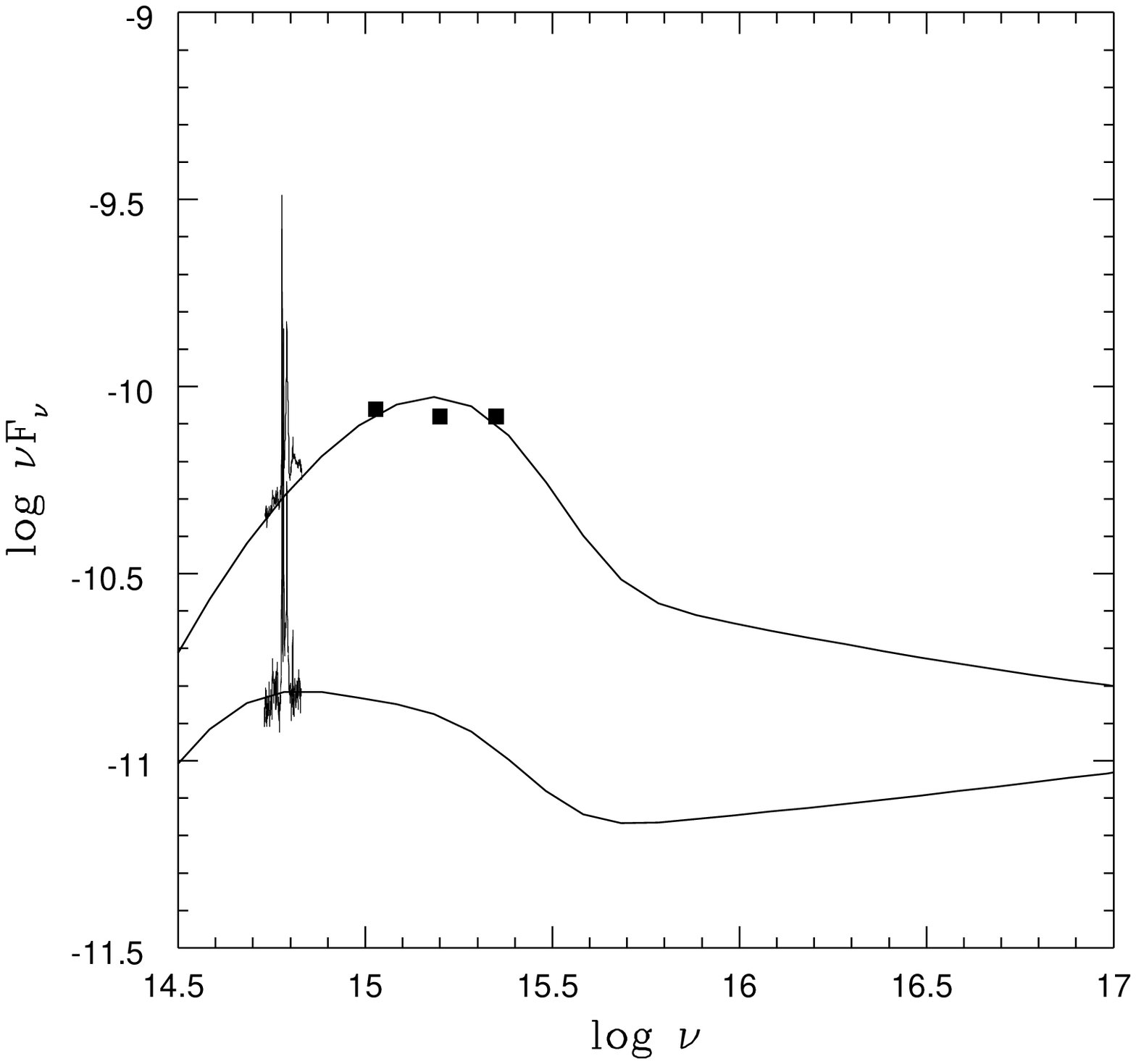}


\epsfbox{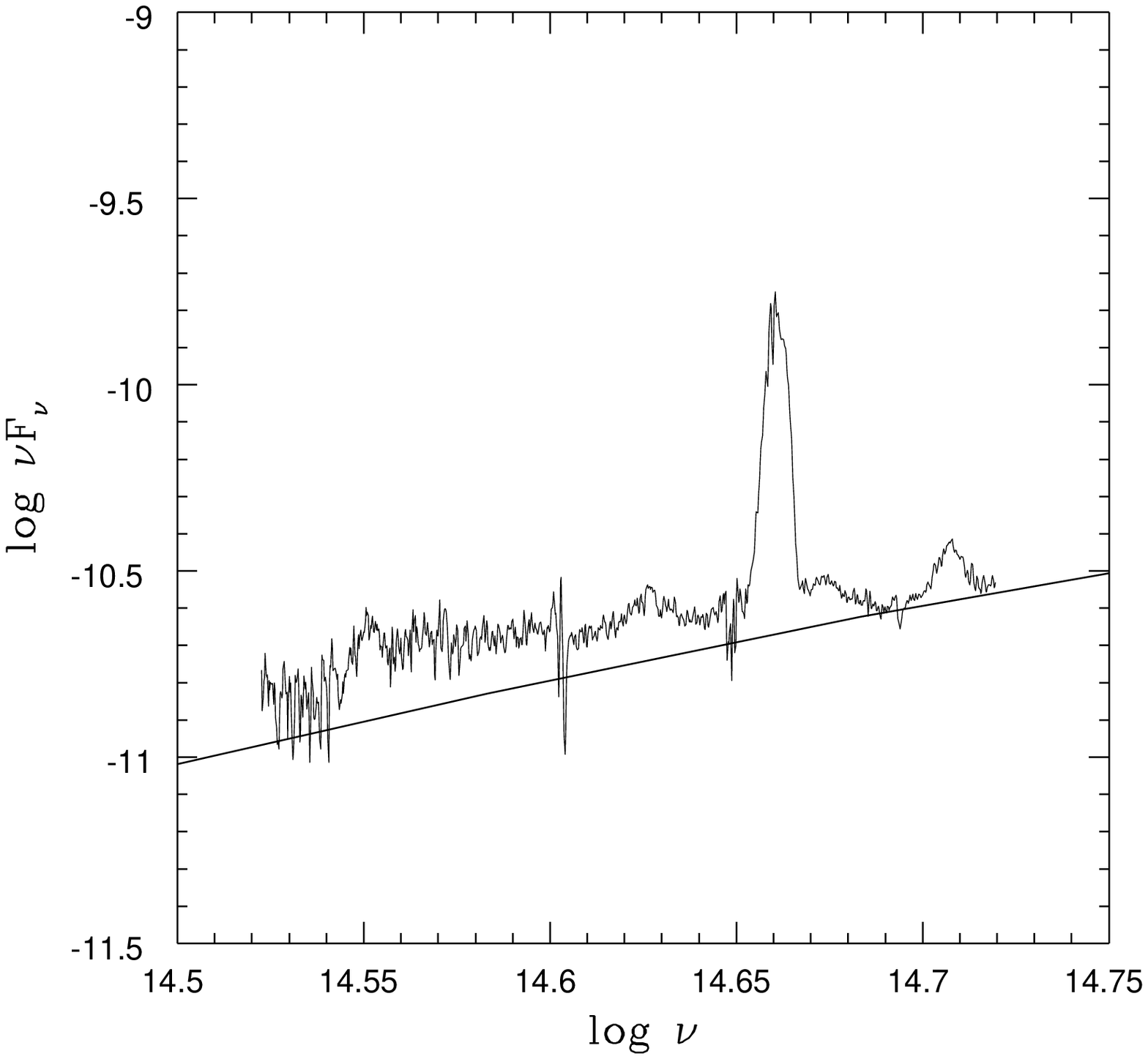}


\epsfbox{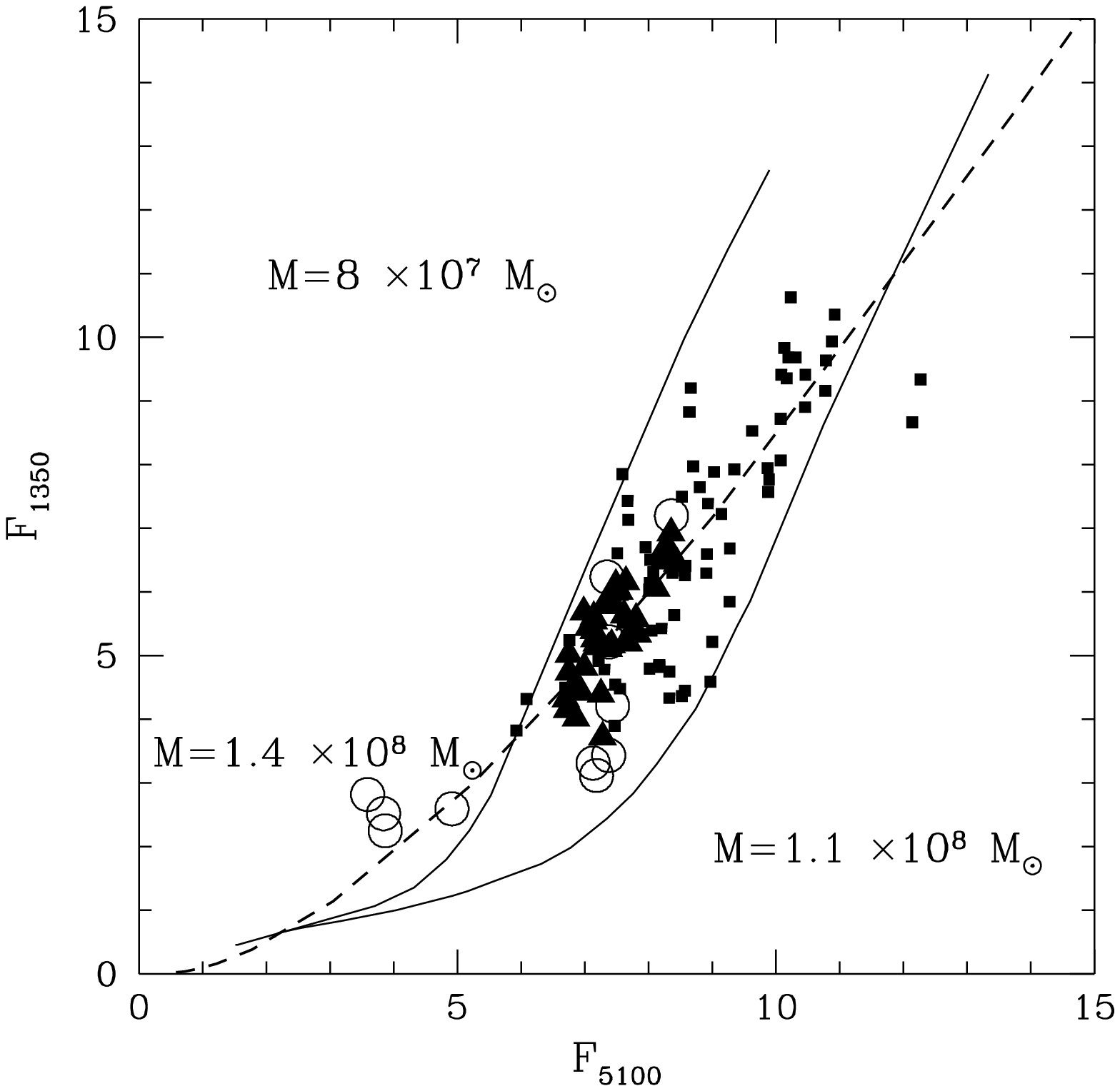}


\epsfbox{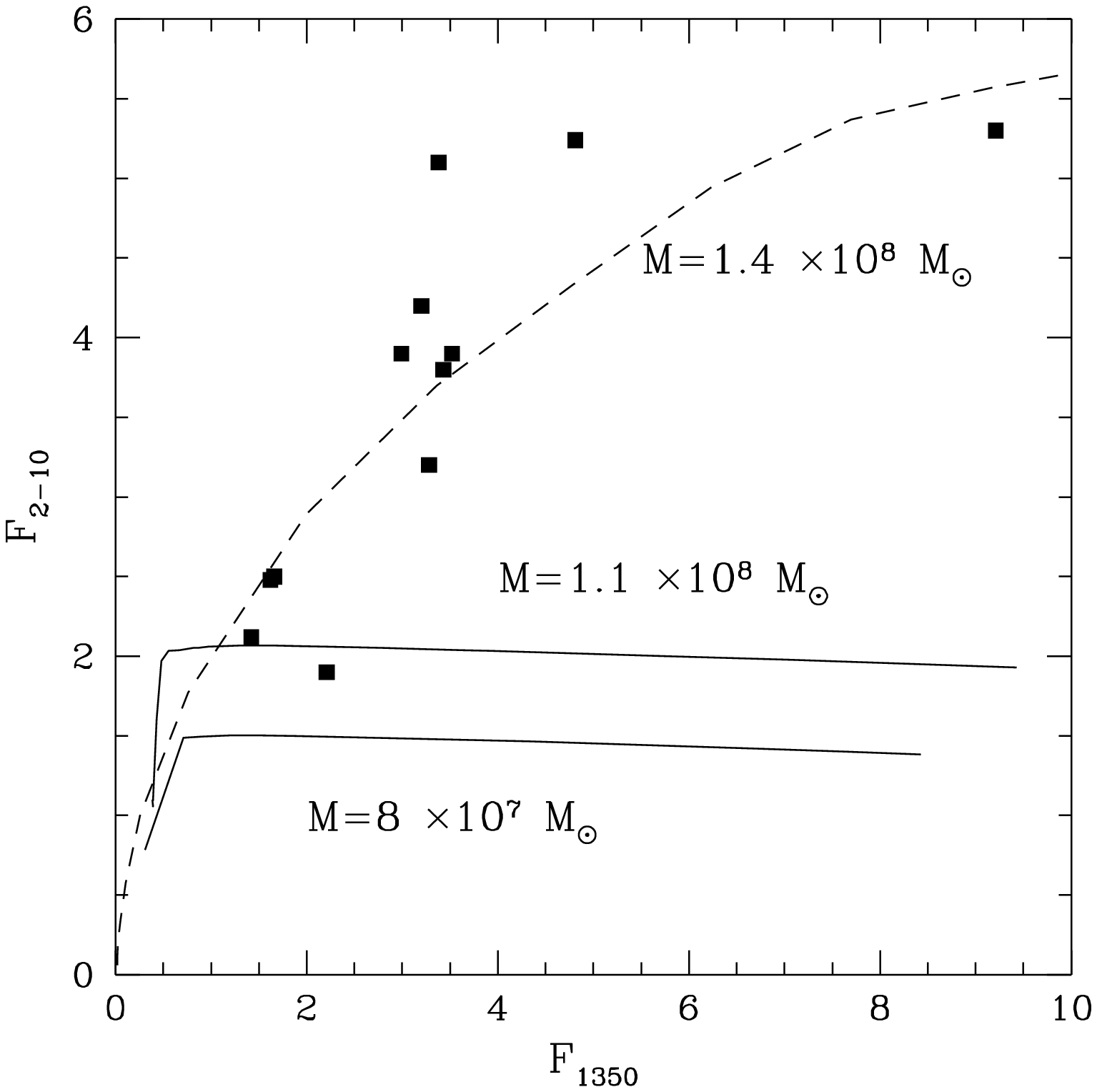}

\end